\documentclass[aps,prb,twocolumn,superscriptaddress,longbibliography]{revtex4-2}
\usepackage{amsmath,amssymb}
\usepackage[pdftex]{hyperref,graphicx}
\hypersetup{colorlinks = true, urlcolor = blue, linkcolor = blue, citecolor = blue}
\usepackage{physics}
\usepackage{xcolor}
\usepackage{bm}

\begin{document}

\title{Strange metallicity of moir\'e twisted bilayer graphene}
\author{Sankar Das Sarma}
\affiliation{Condensed Matter Theory Center and Joint Quantum Institute, Department of Physics, University of Maryland, College Park, Maryland 20742, USA}
\author{Fengcheng Wu}
\affiliation{School of Physics and Technology, Wuhan University, Wuhan 430072, China}
\affiliation{Wuhan Institute of Quantum Technology, Wuhan 430206, China}

\date{\today}

\begin{abstract}
Recent measurements in several different laboratories report the observation of an approximately linear-in-temperature resistivity with a large twist-angle-dependent slope (or temperature coefficient) in moir\'e twisted bilayer graphene (tBLG) down to a few K and sometimes to much lower temperatures.  In this note, we theoretically discuss this `strange metal' linear-in-temperature transport behavior from the perspective of resistive scattering by acoustic phonons, emphasizing the aspects of the transport data, which are and which are not consistent with the phonon scattering mechanism.  Extensive theoretical comparison with a new experiment [A. Jaoui et al., \href{https://www.nature.com/articles/s41567-022-01556-5}{Nature Physics \textbf{18}, 633 (2022)}] is the central new aspect of this work.
\end{abstract}

\maketitle

\section{INTRODUCTION and BACKGROUND}
An impressive recent experiment \cite{jaoui2021quantum} reports extensive transport measurements for the electrical resistivity $\rho (T, n, \theta)$ of tBLG as a function of temperature ($T$), twist angle ($\theta$), and carrier/doping density ($n$).  There are existing works \cite{Polshyn2019,Cao2020PRL} presenting similar tBLG transport results already, but Ref.~\onlinecite{jaoui2021quantum} provides more extensive resistivity data. This work \cite{jaoui2021quantum} has attracted attention in the context of much-discussed `strange metal' behavior in condensed matter physics \cite{Senthil2021}. The key message of Ref.~\onlinecite{jaoui2021quantum} is that the linear-in-$T$ behavior exists over a  large temperature range (at least, for some specific doping induced band fillings), making the behavior consistent with the putative `strange metal' behavior much discussed in the literature on strongly correlated materials \cite{greene2020strange,Zaamen2019,Cha2021PRL,hartnoll2021planckian}.  Although `strange metallicity' is not a sharply defined concept,  with a universally accepted description, it is associated with a linear-in-$T$ resistivity (which is also often, but not always, anomalously large) existing over a substantial temperature range with the underlying scattering mechanism for the linearity not obviously apparent.   For strange metals, therefore, the key question is what causes the large linear-in-$T$ resistivity. In this note, we discuss the possibility that the linear-in-$T$ resistivity observed in Ref.~\onlinecite{jaoui2021quantum} arises (at least partially) from acoustic phonon scattering, using a minimal Dirac model for the tBLG band structure.  This viewpoint has earlier been propounded in our theoretical works \cite{wu2019phonon,sarma2020electron}  in the context of the tBLG experimental works in \cite{Polshyn2019,Cao2020PRL}.  Ref.~\onlinecite{Polshyn2019} indeed concludes that the phonon scattering mechanism is the cause for the linear-in-$T$ dependence of $\rho (T, n, \theta)$ in tBLG whereas Ref.~\onlinecite{Cao2020PRL} suggests that the strangeness in tBLG arises from quantum criticality.  The new work in Ref.~\onlinecite{jaoui2021quantum} claims that quantum criticality is the underlying mechanism for the tBLG strangeness, based mainly on the observation that the ``metallic ground state features a $T$-linear resistivity extending over three decades in temperature, from 40 mK to 20 K, spanning a broad range of dopings including those where a correlation-driven Fermi surface reconstruction occurs"\cite{jaoui2021quantum}. It is often stated in the literature that some unknown hidden quantum criticality may be giving rise to an observed linear-in-$T$ resistivity.  While such a possibility can never be ruled out, no known microscopic itinerant metallic quantum criticality has been shown to decisively lead to any linear-in-$T$ resistivity. Neither quantitative nor qualitative arguments are provided for how and why quantum criticality leads to a linear-in-$T$ resistivity at $T>10$ K, or the nature of the putative quantum critical point leading to strangeness in tBLG.  There is, in fact, no theoretically established and experimentally verified generic mechanism for producing an extended linear-in-$T$ metallic resistivity except for scattering by acoustic phonons.

The fact that acoustic phonon scattering leads to a linear-in-$T$ metallic resistivity for $T>T^*$ , where $T^*$ is a characteristic temperature dependent on both phonon and electron parameters,  has been known since the 1930s. \cite{ziman1972principles} The basic idea is simple: At `high T' the phonons become classical with equipartition, and their thermal occupancy increases linearly with $T$, consequently producing a $T$-linear resistivity.  In regular  3D metals this typically happens for $ T>40$ K or so, which, in the dimensionless electronic temperature units, implies an extremely low $T/T_\text{F}$ of $O (10 ^ {-3})$, where $T_\text{F}$ is the metallic Fermi temperature.  For $ T \ll T^*$, the phonon induced resistivity goes as $T^4$ $(T^5)$ in 2D (3D) metals, and is extremely small.  The crossover temperature scale as $T^* \sim T_{\text{BG}}/6$ or $T_D/6$, depending on whether $T_{\text{BG}}$ or $T_{\text{D}}$ is smaller.  The Bloch--Gr\"uneisen (BG) temperature is given by
\begin{equation}
    T_{\text{BG}} = 2\hbar k_{\text{F}}  v_{\text{ph}},
    \label{eq:T_BG}
\end{equation}
where $k_{\text{F}}$ is the electronic Fermi momentum and $v_{\text{ph}}$ is the acoustic phonon (i.e. longitudinal sound) velocity.  The Debye temperature $T_{\text{D}}$ of most materials is around a few hundred kelvins, and in metals, by virtue of the very large $k_{\text{F}}$, $T_{\text{BG}} \gg T_{\text{D}}$, so $T^* \sim T_{\text{D}}/6 \sim 40-50$ K.  It has been argued \cite{Hwang2019PRB} that many 2D strongly correlated materials tend to have very low carrier density, and hence $k_{\text{F}} \sim n^{1/2}$ is small, leading to rather low values of $T^* \sim T_{\text{BG}}/6 \sim 1-10 $ K (or even lower depending on the actual doping density), since for low-density systems $T_{\text{D}} \gg T_{\text{BG}}$.  Thus, in principle, phonon scattering could lead to strange metallicity with a linear-in-$T$ resistivity going down to very low temperatures in dilute systems, but whether it actually happens in any of the actively-studied strange metals remains an open question.  One thing is certain;  if the linear-in-$T$ behavior survives to arbitrarily low $T$, it is unlikely to arise entirely from phonon scattering since the resistive scattering by phonons must be strongly suppressed for $T \ll T^*$, where quantum degeneracy strongly suppresses the phonon thermal occupancy, changing the $T$-linear resistivity to an almost unobservable $T^4$ dependence (in 2D) with a complicated crossover regime (for $T<T^*$) in between the two different power law behaviors.

Since scattering by acoustic phonons is the only known generic mechanism for producing strange metallicity, i.e., a linear-in-$T$ resistivity over a large range of temperatures (but not extending to arbitrarily low temperatures), it behooves us to critically discuss the new experimental findings of Ref.~\onlinecite{jaoui2021quantum} using phonon scattering within a minimal flat band Dirac model (so that there is a minimum number of unknown free parameters) as the underlying mechanism, in order to discern what aspects of the data are consistent with the phonon mechanism and where new physics may be lurking.  This is what we do in the current paper. We note that the current work is an expanded and enhanced application of our earlier  transport theories on electron-phonon interaction induced temperature dependence of graphene and tBLG resistivity. \cite{Hwang2008PRB,wu2019phonon}  The new aspects of the current work are applying the theory quantitatively to new experimental data over a much broader temperature and density range than before so as to critically examine where new theoretical thinking might be necessary.

\section{THEORY and RESULTS}

Now, we focus on tBLG, where $T_{\text{BG}}<T_{\text{D}}$, for all doping densities, so $T^* \sim T_{\text{BG}}/6 \sim n^{1/2}$, where $n$ is the effective doping density.  We note that in graphene $T_{\text{BG}}$ tends toward zero as the Dirac point or the charge neutrality point (CNP) is approached since the effective doping vanishes at the CNP.  Thus, indeed, as a matter of principle, phonons could produce a $T$-linear resistivity in graphene for arbitrarily low $T$ and for arbitrarily low carrier density.  Of course, this linear-in-$T$ phonon induced resistivity may be overwhelmed, particularly at very low temperatures, by other resistive contributions (e.g. disorder and impurity scattering) and thus become unobservable. Also, such a phonon-induced linear-in-$T$ resistive behavior, even if it exists, can only happen at very small doping near the CNP, and cannot explain a very low-$T$ linearity at large doping values with finite $k_{\text{F}}$.

\begin{figure}[t]
    \includegraphics[width=1\columnwidth]{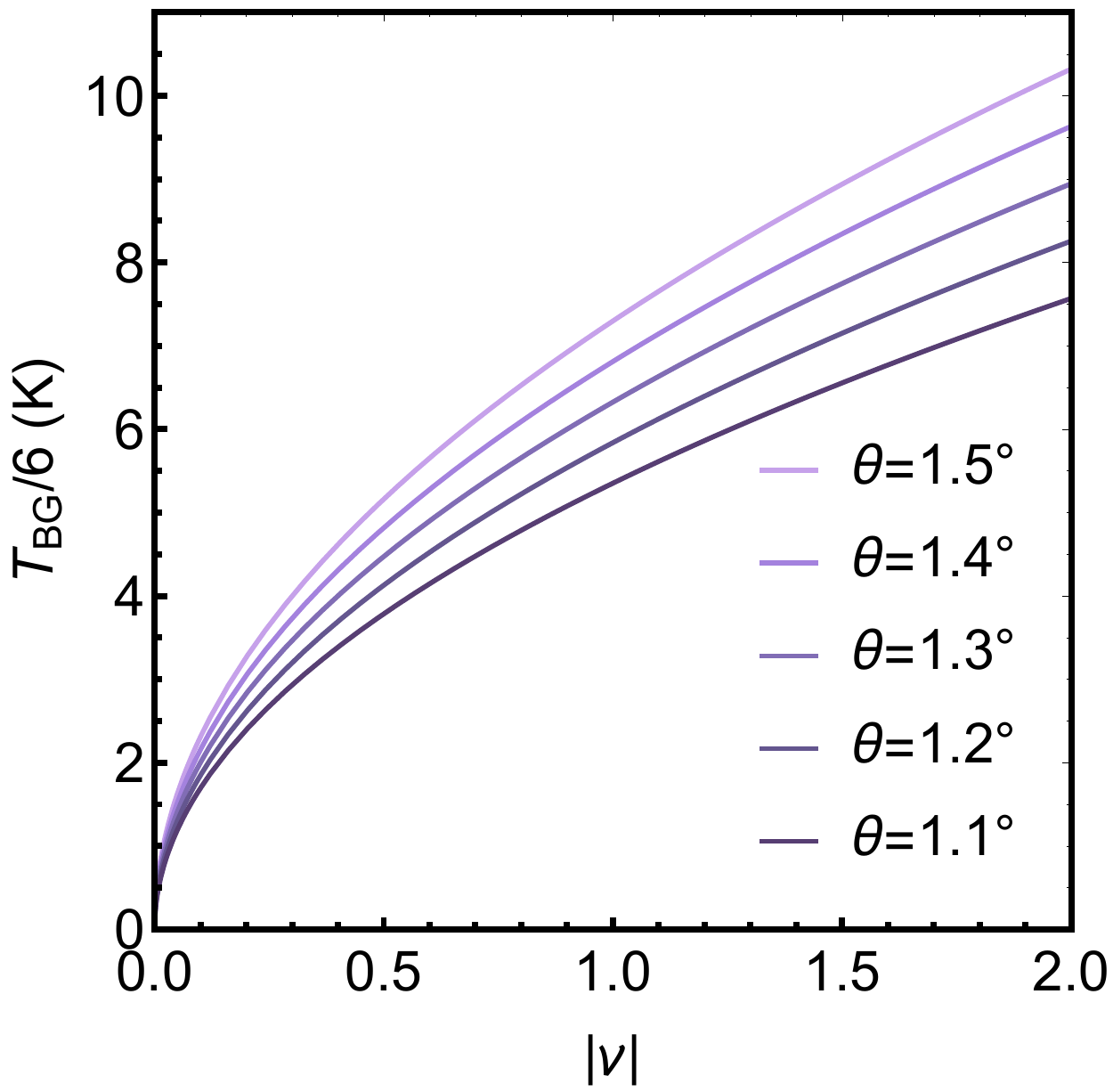}
    \caption{The linear-in-$T$ resistivity onset temperature $T^* \sim T_{\text{BG}}/6$ as a function of the moir\'e band filling factor $\nu$ for different twist angles.}
    \label{fig:1}
\end{figure}

In Fig.~\ref{fig:1}, we show our calculated $T^*$ for tBLG, for a few values of the twist angle, as a function of doping expressed in terms of the band filling $\nu$, where $\nu = 0  (4)$ indicates the unfilled empty moir\'e conduction band with the Fermi level at the CNP (the filled moir\'e conduction band with the Fermi level at the band edge).  We show our results in terms of the band filling $\nu$ rather than the carrier density $n$ in order to be consistent with the data presentation in Ref.~\onlinecite{jaoui2021quantum}.  Note that Fig.~\ref{fig:1} stops at $\nu = 2$ simply because the Dirac approximation is manifestly invalid for large band filling beyond the van Hove singularities in the moir\'e band structure.   In fact, our minimal model applies only for $\nu<2$.

We note that, as expected $T^*$,  increases (decreases) with increasing (decreasing) band filling, simply because a higher filling implies, within our minimal 2D Dirac model, a higher carrier density and consequently a higher $k_{\text{F}}$ in Eq.~\eqref{eq:T_BG}. The variation with $\theta$ arises from the moir\'e band structure effect included in our theory within the minimal Dirac model.  We note that our minimal model predicts that for $ T>10$ K and for $\theta = 1.1^\circ-1.5^\circ$  (the experimental range for Ref.~\onlinecite{jaoui2021quantum}), the linear-in-$T$ behavior arising from phonon scattering should prevail everywhere for Ref.~\onlinecite{jaoui2021quantum} measurements, and we therefore focus on the resistivity behavior for $T>10$ K, leaving the discussion of the intriguing low-$T$ transport behavior toward the end of this paper.

\begin{figure}[t]
    \includegraphics[width=1\columnwidth]{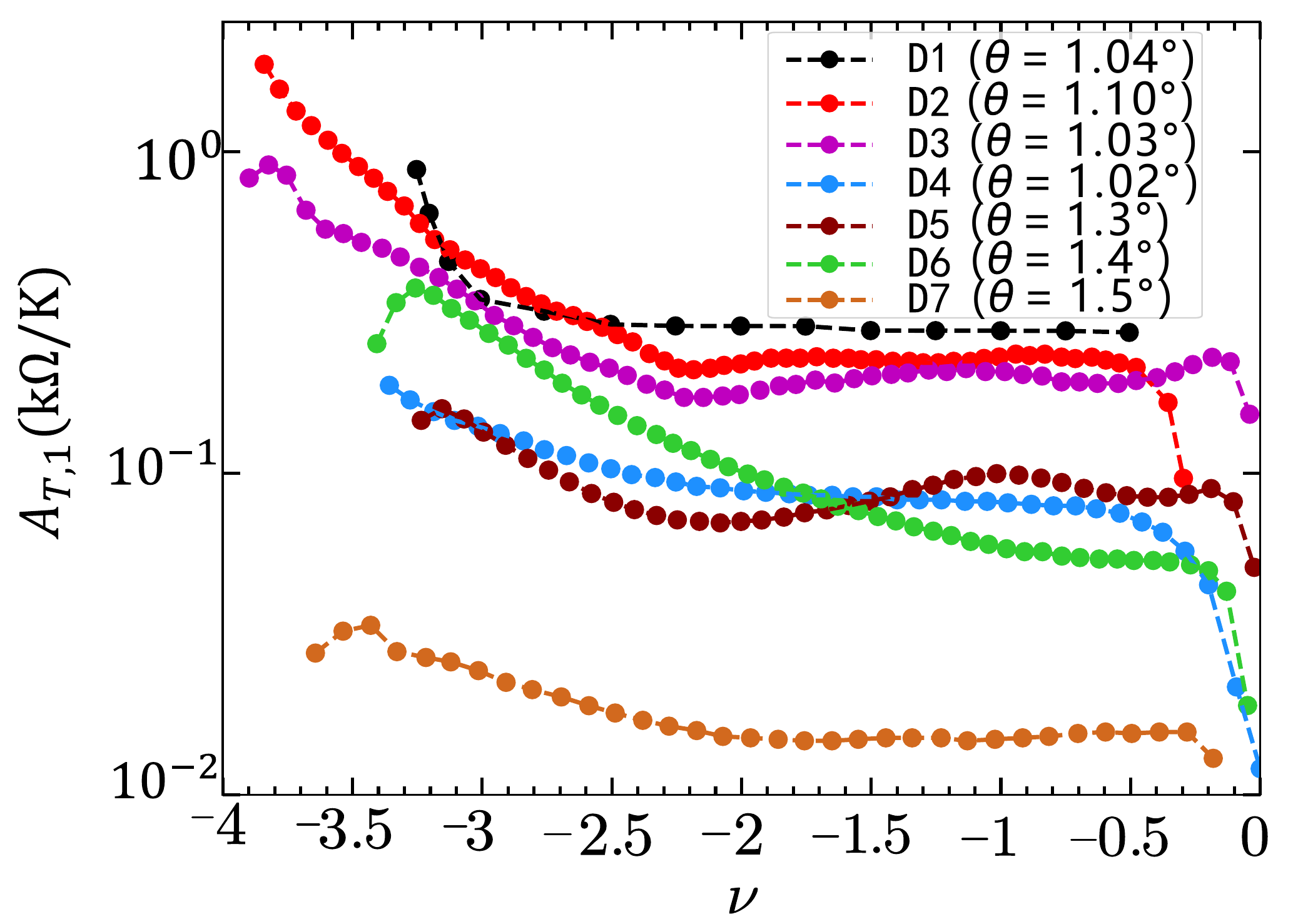}
    \caption{The linear-in-$T$ resistivity slope $A_{T,1}=d\rho/dT$ for $T>10$ K as a function of the filling factor $\nu$ in devices with different twist angles. The data are obtained from Ref.~\onlinecite{jaoui2021quantum}  }
    \label{fig:2}
\end{figure}

In Fig.~\ref{fig:2}, we show the filling- and twist-angle- dependent linear temperature coefficient $A_{T,1} = d\rho(T)/d T$ for all the experimental data  in Ref.~\onlinecite{jaoui2021quantum}, obtained for $T>10$ K. \cite{jaoui2022}   We emphasize that Fig.~\ref{fig:2}  is simply a succinct summary (for our purpose) of the resistivity experimental data from Ref.~\onlinecite{jaoui2021quantum}, shown as a temperature derivative of the resistivity plotted as a function of both density and twist angle.\cite{jaoui2022} Any theory trying to explain the `strange metallicity' reported in Ref.~\onlinecite{jaoui2021quantum} must at least be able to make sense of this `higher-temperature' $T>10\text{ K}$ data.  The data for all seven samples of Ref.~\onlinecite{jaoui2021quantum} are shown in Fig.~\ref{fig:2} corresponding to seven different values of the twist angle between $1^\circ$ and $1.5^\circ$. We first discuss several salient features of Fig.~\ref{fig:2} before comparing our phonon theory with the data.

First, the density (angle) dependence of $A_{T,1}$ is weak (strong).  This is absolutely obvious in Fig.~\ref{fig:2}, but not so apparent in Ref.~\onlinecite{jaoui2021quantum}, where $\rho (T)$ for many values of $\nu$ are shown displaced in the same figure, and different $\theta$ values corresponding to different samples are plotted in different figures, making a comprehensive understanding of the density/angle dependence a challenge.  The Fig.~\ref{fig:2}, which is a direct replot of the data in Ref.~\onlinecite{jaoui2021quantum} (and kindly supplied to us by the experimentalists themselves \cite{jaoui2022}), makes it clear that the temperature derivative of resistivity, i.e., the slope of  $\rho (T)$ with respect to $T$, is almost independent of density, but very strongly dependent on the twist angle, at least for all the high-$T$ ($>10$ K) data.  This qualitative behavior (i.e. weak/strong dependence on $\nu/\theta$) is easily explainable based on the phonon scattering mechanism, as discussed below and in our earlier work. \cite{wu2019phonon,sarma2020electron} One should remember that each twist angle in Fig.~\ref{fig:2} corresponds to a different sample, with likely unknown variations arising from disorder \cite{Hwang2020PRR} and twist angle fluctuations \cite{Wilson2020PRR}, making a direct quantitative comparison between results at different angles difficult (and perhaps also explaining some of the nonmonotonicity and the crossing of various lines for $|\nu|>1$).  But the overall trend is clear: As the twist angle decreases from $1.5^{\circ}$ to $1.1^{\circ}$, the resistivity coefficient $A_{T,1}$  is enhanced by more than a factor of 10 at, for example,  $\nu=0.5$ in Fig.~\ref{fig:2}.  By contrast, $A_{T,1}$ remains essentially a constant as $\nu$ varies from 0.3 to 2 at the twist angle $1.5^{\circ}$ and from 0.5 to 2.3 at the angle of $1.1^{\circ}$.  Any theory must explain this dichotomy in the temperature coefficient of the measured tBLG resistivity with respect to its density and twist angle dependence. Phonons can do this easily and naturally as discussed below. (We note that the drop-off at very low filling  for all samples in Fig.~\ref{fig:2} happens close to the CNP, where the carrier density is very small, and many other effects, including thermal carrier excitation and disorder, come into play. In addition, the CNP in samples of of Ref.~\onlinecite{jaoui2021quantum} can have a small substrate-induced gap, which makes the temperature dependent resistivity near CNP to be insulating in contrast to the metallic behavior of interest \cite{jaoui2022}.)

The minimal theory for acoustic phonon scattering induced carrier resistivity for moir\'e Dirac carriers gives the following \cite{wu2019phonon} formula
\begin{equation}
\begin{aligned}
\rho &= \frac{ 32 F(\theta) D^2 k_{\text{F}}}{ g_s g_v g_l e^2  \rho_m  v_{\text{F}}^{*2} v_{\text{ph}}} I(\frac{T}{T_{\text{BG}}}),\\
I(z)& = \frac{1}{z} \int_0^1 dx x^4 \sqrt{1-x^2} \frac{e^{x/z}}{(e^{x/z}-1)^2},
\end{aligned}
\label{eq:rho_TBG}
\end{equation}
where $D$ is the deformation potential electron-phonon coupling, $v_{\text{F}}^*$ is the Fermi velocity in the moir\'e band, $g_{s, v, l}$ are the degeneracy factors (all equal to 2), and $\rho_m$ is the atomic mass density of graphene.  The function $F(\theta)$ is a weakly varying function of $\theta$, dependent on the detailed moir\'e band structure, which is discussed in detail in Ref.~\onlinecite{wu2019phonon}.  Typically, $0.5<F(\theta)<1$.  Since our interest is in the equipartition regime of $T>T^*$, where the linear-in-$T$ resistivity manifests, we can obtain from Eq.~\eqref{eq:rho_TBG}, for $T>T^*$, an approximate expression for $A_{T,1} = d\rho/dT$ as
\begin{equation}
    A_{T,1} = d\rho/dT \propto \Big[ \frac{1}{v_{\text{F}}^{*2}} \Big(\frac{D}{v_{\text{ph}}}\Big)^2\Big] T.
\label{eq:AT1}    
\end{equation}
Thus the main features of the phonon-induced linear-in-$T$ resistivity are: (1) it has no dependence on $k_{\text{F}}$, and hence on carrier density or band filling, and (2) it depends strongly on the moir\'e band velocity $v_{\text{F}}^{*}$ through the $1/v_{\text{F}}^{*2}$ dependence -- as the moir\'e band flattens approaching the magic angle, the flatband-renormalized Fermi velocity $v_{\text{F}}^*$ is strongly suppressed, enhancing the phonon-induced resistivity slope by a factor of $v_{\text{F}}^{*-2}$.  Thus, the two most significant qualitative features of the data presented in Ref.~\onlinecite{jaoui2021quantum} are completely explained by phonon scattering: {\it Weak} dependence on carrier density or band filling and {\it strong} dependence on the twist angle through the $\theta$ dependence of $v_{\text{F}}^*$.   We know of no other theoretical model for tBLG transport, which can naturally explain these two key qualitative features of the tBLG strange metallicity.  We also mention that Eq.~\eqref{eq:AT1} for the temperature coefficient of the linear-in-$T$ tBLG resistivity is similar to the corresponding result for the untwisted monolayer graphene (MLG) \cite{Hwang2008PRB}, which has been experimentally verified \cite{Efetov2010}, except for the appearance of the moir\'e flat-band Fermi velocity $v_{\text{F}}^*$ in Eq.~\eqref{eq:AT1} whereas for regular untwisted graphene, the Fermi velocity is that for regular MLG, $v_{\text{F}}>v_{\text{F}}^*$.

\begin{figure}[t]
    \includegraphics[width=1\columnwidth]{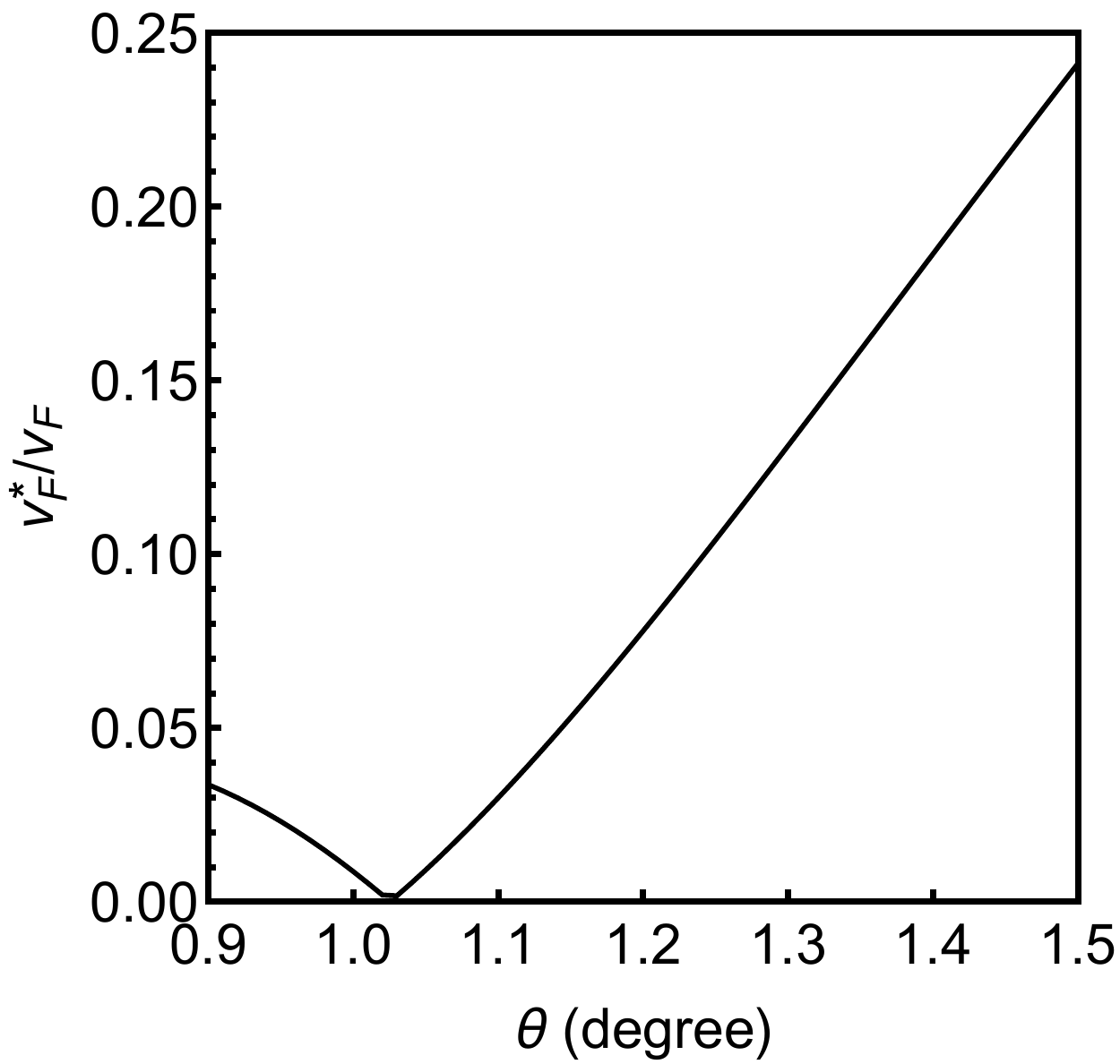}
    \caption{The twist angle dependence of $v_\text{F}^*/v_\text{F}$ obtained from the BM model. $v_\text{F}$ and $v_\text{F}^*$ are the Fermi velocity at the Dirac points in MLG and tBLG, respectively.}
    \label{fig:3}
\end{figure}

To show the obvious large effect of the flat band suppression of the Fermi velocity in the tBLG compared with untwisted MLG, we plot in Fig.~\ref{fig:3} the calculated ratio $v_{\text{F}}^*/v_{\text{F}}$ as a function of the twist angle, obtained using the standard Bistritzer-MacDonald (BM) band structure model for the moir\'e system.\cite{Bistritzer2011}  Within the BM model, $v_{\text{F}}^*$ vanishes at the magic angle $\sim 1^{\circ}$, and is as small as $v_{\text{F}}/4$ even for a large twist angle of $1.5^{\circ}$, which corresponds to a factor of 16 enhancement in the tBLG resistivity and its temperature coefficient compared with that in the untwisted MLG.

We use the full Eq.~\eqref{eq:rho_TBG} to calculate the theoretical $d\rho/dT$ using the BM moir\'e band structure \cite{Bistritzer2011} to compare with the experimental data shown in Fig.~\ref{fig:2}.  The only unknown, which we use as an adjustable parameter, is the deformation potential coupling $D$ in tBLG, whose value we fix by demanding agreement between theory and experiment at the largest twist angle $\theta=1.5^{\circ}$ (and then keep it fixed throughout).  Note that $D$ just determines the overall scale of the resistivity, not its functional dependence on carrier density and twist angle.  Thus, the qualitative finding of a weak (strong) density (angle) dependence is generic in the phonon theory and independent of the value of $D$.  We use the accepted graphene values for the sound velocity and mass density for the transport calculation, $\rho_m=7.6\times 10^{-8}$ g/cm${^2}$ and $v_{\text{ph}}=2\times 10^6$ cm/s. In Eqs.~\eqref{eq:rho_TBG} and \eqref{eq:AT1} for the resistivity of tBLG, the Fermi velocity $v_{\text{F}}^*$ takes the renormalized value due to moiré superlattices, but the mass density and sound velocity take the same values as those in monolayer graphene. The sound velocity of longitudinal acoustic phonons in tBLG is found to be not significantly altered compared to that in monolayer graphene, as shown in Ref.~\onlinecite{koshino2019moire}. Therefore, we take $v_{\text{ph}}$ to be the sound velocity of monolayer graphene.  $\rho_m$ is taken to be the mass density of monolayer graphene in Eq.~\eqref{eq:rho_TBG}, while the additional layer degree of freedom in tBLG is taken into account by the factor $g_{l}$. A detailed description of electron-phonon coupling in tBLG can be found in Ref.~\onlinecite{wu2019phonon}.

\begin{figure}[t]
    \includegraphics[width=1\columnwidth]{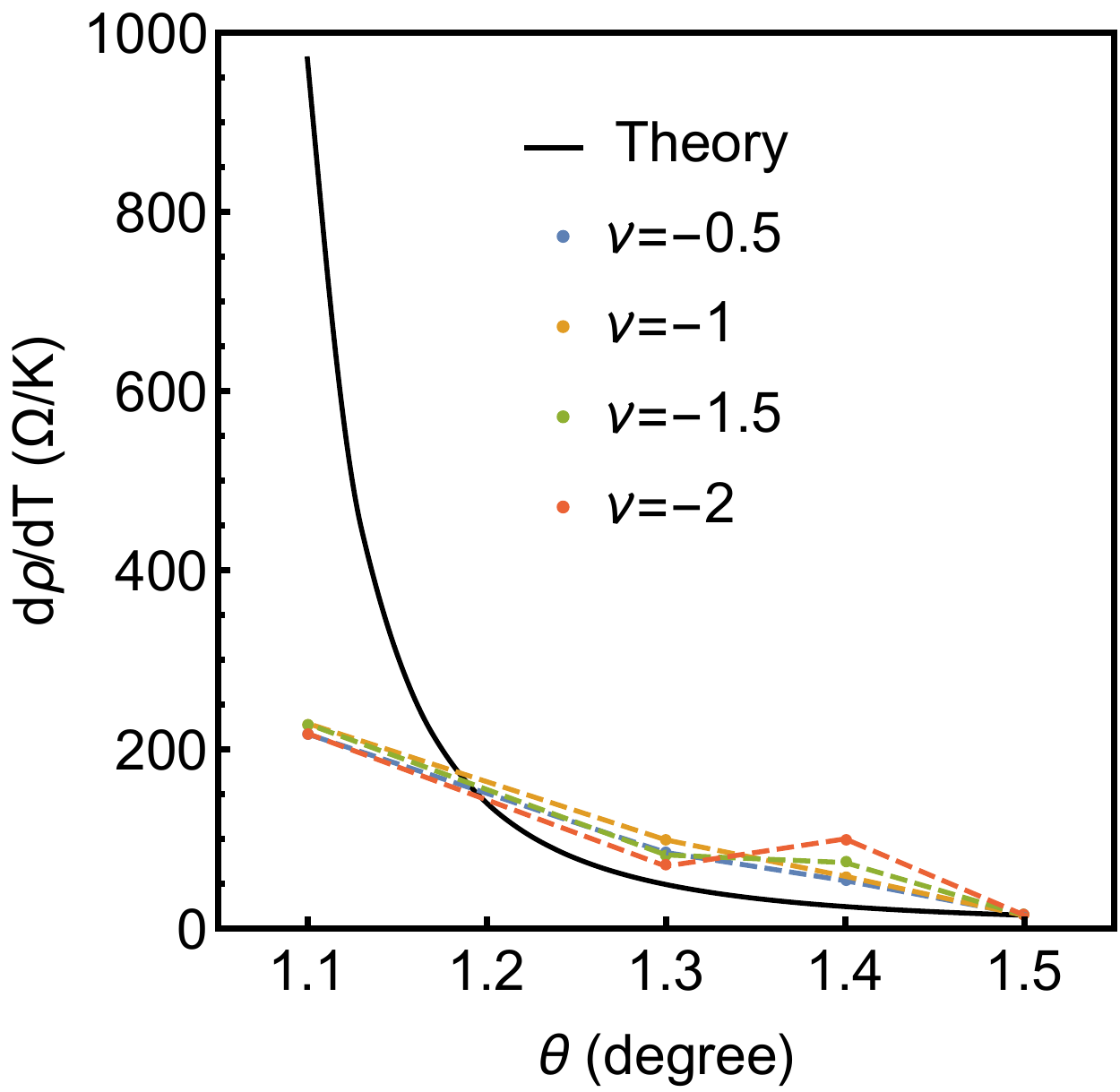}
    \caption{Comparison between theoretical (black curve) and experimental (colored dots) slope $d\rho/d T$ for $T>T^*$. The experimental data are replotted from Fig.~\ref{fig:2}. The theoretical curve is calculated based on Eq.~\eqref{eq:rho_TBG} with a fixed $D=100$ eV.}
    \label{fig:4}
\end{figure}

In Fig.~\ref{fig:4}, we provide a direct comparison of the theoretical results for the calculated temperature coefficient $A_{T,1} = d\rho/dT$ as a function of the twist angle with the corresponding experimental results from Ref.~\onlinecite{jaoui2021quantum} as shown in Fig.~\ref{fig:2}.  The experimental data for different band fillings (i.e. different densities) are replotted from Fig.~\ref{fig:2} for clarity, and the solid line is the theoretical calculation within the minimal Dirac model using the BM band structure.

The deformation potential fit at $\theta = 1.5^{\circ}$ gives $ D = 100$ eV, which is what we use throughout in Fig.~\ref{fig:4}.  It is reassuring that the same value of $D$ ($\sim 80-100$ eV) is necessary here for agreement with the tBLG data of Ref.~\onlinecite{jaoui2021quantum} as what was needed \cite{wu2019phonon} to obtain quantitative agreement with the earlier tBLG transport results in Refs.~\onlinecite{Polshyn2019} and \onlinecite{Cao2020PRL}.  For regular MLG, the quoted value for $D$ ranges between 20-40 eV \cite{wu2019phonon,Hwang2008PRB}, and for tBLG we need roughly a $D$ which is 2-3 times larger.  It is possible that the twisted graphene layers indeed have larger values of deformation potential coupling than regular untwisted graphene, but for us $D$ is simply one free parameter of the theory, which we fix at the largest twist angle, for which Ref.~\onlinecite{jaoui2021quantum} reports its experimental results.  It may be useful to point out in this context that, in general, the deformation potential coupling is often unknown in materials, and different ways of estimating the coupling constant often differ by factors of 2-3.  For example, even in extensively studied GaAs, the deformation potential coupling differs by a factor of 2 between optical and transport measurements, and typically its precise value for calculating the resistivity of doped GaAs has to be fixed by comparing with the transport data.\cite{Kawamura1990}  It is thus quite possible for the moir\'e structure to modify the deformation potential coupling by a factor of 2-4 in the tBLG system.  More work should be carried out to definitively settle this question. In fact, it has been argued \cite{ishizuka2021purcell} that an effective ``Purcell effect'' associated with the compression of the Wannier orbitals in tBLG strongly enhances the electron-phonon coupling, which is certainly consistent with our finding of an effective enhanced deformation potential necessary for quantitatively explaining the tBLG resistivity data.  

Another possibility is that the sound velocity, $v_{\text{ph}}$, is somehow suppressed by the moir\'e superlattices, decreasing it from its nominal graphene value of $2\times 10^6$ cm/s. Since $A_{T,1} \propto (D/v_{\text{ph}}) ^2$, a decrease of $v_{\text{ph}}$ is equivalent to an increase in the deformation potential constant compared with its nominal MLG value.  It is also possible that both $D$ and $v_{\text{ph}}$ are affected by the moir\'e superlattices, leading to an enhancement of the effective coupling.  These are all possibilities that future work should explore, but for our purpose, it suffices to take the overall scale as a phenomenological tuning parameter and fix it by comparing with the experiment at the highest available twist angle of 1.5$^{\circ}$ in Ref.~\onlinecite{jaoui2021quantum}.  We note that a suppressed $v_{\text{ph}}$ in the twisted moir\'e system would also decrease the crossover temperature scale $T^*$ since $T_{\text{BG}} \propto v_{\text{ph}}$, leading to the linear-in-$T$ resistivity extending to lower temperature scales of the order of 1 K or less.
 Our calculation assumes that the phonons are unaffected by moir\'e superlattice and retain their pristine values for MLG.  It is, however, possible that the twisting strongly affects the acoustic phonons through strain and other effects.  If so, $D/v_{\text{ph}}$ could increase considerably from their MLG value, providing a unified explanation for why the resistivity is linear down to low-T and why it is so strongly enhanced.
We do not, however, pursue such a line of argument further in this paper since our goal is not seeking a precise quantitative agreement with experiments, but to understand which aspects of the data in Ref.~\onlinecite{jaoui2021quantum} are generically and qualitatively in agreement with the phonon scattering mechanism so that we can in the future focus on the unknown mechanisms perhaps in play here beyond phonon scattering.

The results presented in Fig.~\ref{fig:4} show that the phonon mechanism explains the experimental data of Ref.~\onlinecite{jaoui2021quantum} for all samples for twist angle down to $\sim 1.17^{\circ}$, but then the theory predicts much larger resistivity than the experiment.  By fine-tuning the value of $D$ to obtain the best recursive fit to the experimental data may extend the regime of agreement down to $\sim 1.15^{\circ}$, but we do not believe that this is a useful exercise as the experiment and theory most definitely disagree at low enough twist angle somewhere just below $1.2^{\circ}$.  A trivial possibility, which we do not consider, is that $D$ somehow decreases for $\theta<1.2^{\circ}$, for example, experiment and theory would agree well at $\theta=1.1^{\circ}$ if we arbitrarily reduce $D$ (at $1.1^{\circ}$ twist angle) to 30 eV.  However, this is just data fitting with no justification.

\begin{figure}[t]
    \includegraphics[width=1\columnwidth]{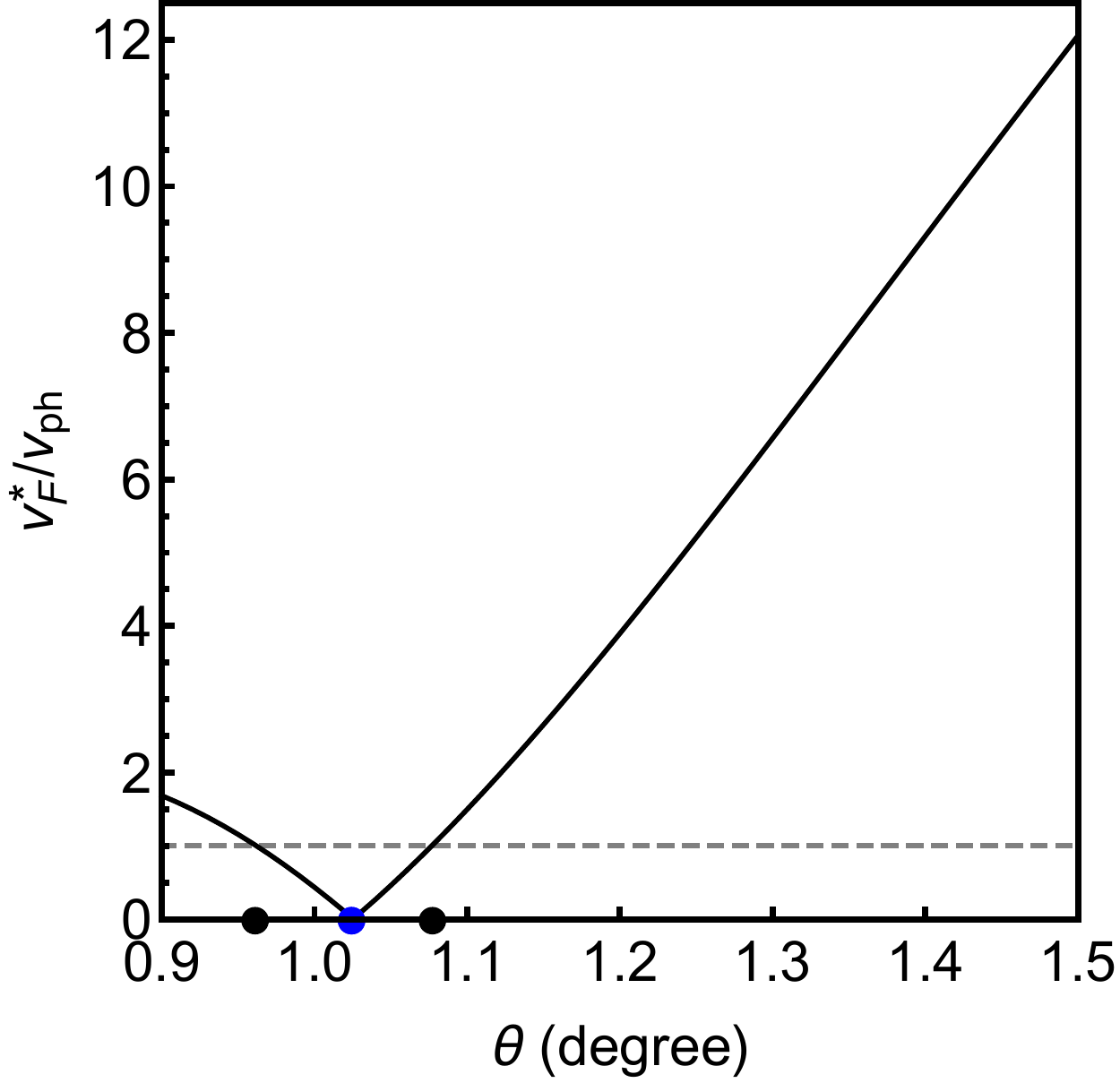}
    \caption{The twist angle dependence of $v_\text{F}^*/v_\text{ph}$ obtained from the BM model. $v_\text{F}^*$ is calculated using the BM model, and $v_\text{ph}$ is assumed to be independent of the twist angle in the calculation. The blue dot marks the magic angle at which $v_\text{F}^*$ vanishes. The two black dots indicate the angles where $v_\text{F}^*=v_{\text{ph}}$.}
    \label{fig:5}
\end{figure}

We believe that the systematic quantitative failure of our transport theory at lower twist angle, apparent in Fig.~\ref{fig:4}, arises from two reasons: (1) as the moir\'e band flattens with decreasing angle, our Dirac approximation becomes increasingly inaccurate quantitatively, and perhaps more importantly, (2) for small twist angles the moir\'e band Fermi velocity $v_{\text{F}}^*$ becomes comparable to the phonon velocity $v_{\text{ph}}$, leading to the inapplicability of the Migdal theorem \cite{migdal1958interaction}, which is the basis of our minimal Boltzmann transport theory.  To demonstrate this point more concretely, we show in Fig.~\ref{fig:5} the calculated ratio $v_{\text{F}}^*/v_{\text{ph}}$ as a function of the twist angle.  We emphasize that for regular MLG, i.e. for untwisted graphene (or equivalently, for very large twist angles), this ratio goes to the very high value of 50 since $v_\text{ph} \sim 2\times 10^6$ cm/s and $v_\text{F} \sim 10^8$ cm/s for regular graphene, and the Migdal approximation is essentially exact.\cite{Roy2014PRB}  But with decreasing $v_{\text{F}}^*/v_{\text{ph}}$ in tBLG, as shown in Fig.~\ref{fig:5}, the Migdal approximation, and hence the leading-order transport theory becomes increasingly inaccurate.\cite{Prange1964} In Fig.~\ref{fig:5}, we denote, on the abscissa,  the points  where $v_{\text{F}}^*=v_{\text{ph}}$ as well as the `magic angle' where the moir\'e band is perfectly flat within the BM approximation producing $v_{\text{F}}^*=0$.  Since $v_{\text{F}}^* \sim v_{\text{ph}}  $ already at $\theta \sim 1.1^{\circ}$, we expect our transport approximation to work only for $\theta>1.1^{\circ}$.  If we arbitrarily set the lower limit of the Migdal approximation to be $v_{\text{F}}^*>3 v_{\text{ph}}  $, then the theory should give reasonable quantitative results for $\theta>1.15^{\circ}$, which is approximately consistent with Fig.~\ref{fig:4}.  We emphasize that there is no controlled way to go beyond the minimal theory when the Migdal approximation no longer applies since all diagrams in the electron-phonon coupling to all orders contribute to the resistivity in a hopelessly strong coupling manner since the electron-phonon coupling itself is also strong by virtue of the small Fermi velocity $v_{\text{F}}^*$ at small angles, exactly where the Migdal approximation also breaks down.

\section{DISCUSSION and CONCLUSION}

We have established a reasonable case that the tBLG linear-in-$T$ resistivity data of Ref.~\onlinecite{jaoui2021quantum} are well-explained by acoustic phonon scattering at higher temperatures ($>10$ K) and at higher twist angles ($>1.15^{\circ} $), suggesting the failure of the Migdal theorem as the reason for the quantitative failure of our theory at lower twist angles.  In addition to providing a reasonable quantitative description of the data \cite{jaoui2021quantum}, the theory has the great intrinsic advantage of providing a natural {\it qualitative} explanation for the weak density (i.e. band filling) and the strong twist angle dependence of the temperature coefficient of the observed linear-in-$T$ resistivity.  We know of no alternative theory which is capable of providing a qualitative understanding of the experimentally observed ( see Fig.~\ref{fig:2} and also the results in Ref.~\onlinecite{jaoui2021quantum}) weak (strong) density (twist angle) dependence of the resistivity as the phonon theory does naturally.

The open question we have not discussed yet is the low-temperature behavior of the resistivity, below 10 K, and actually even below 1 K. We emphasize a key aspect of the experimental data which all theories must address.  The slope of $d\rho/dT$ in the observed linear-in-$T$ resistivity can remain constant throughout the low-$T$ and high-$T$ regimes for certain density ranges at magic angle devices, hinting at one underlying universal transport mechanism in play since it is unlikely that two independent mechanisms would produce the same slope.  In our phonon theory, this constant slope is naturally explained as arising from the phonon equipartition physics coming from the high-$T$ regime.  Any quantum critical or other competing theories must address why the resistivity remains linear to very high $T$ ($\sim$ 50 K) without changing the slope since quantum criticality is a $T=0$ phenomenon. We emphasize that our focus has been on the experimental resistivity of Ref.~\onlinecite{jaoui2021quantum} for $T>10$ K, where the linear-in-$T$ resistivity is fairly generic in all samples essentially for most values of band filling.  Our Fig.~\ref{fig:2} is a precise replotting of the experimental data  for $T>10$ K in Ref.~\onlinecite{jaoui2021quantum}, shown as $d\rho/dT$ for different $\nu$ and $\theta$.  Unfortunately, the experimental data of Ref.~\onlinecite{jaoui2021quantum} at low temperatures  do not reflect a universal behavior, with the temperature dependence of $\rho (T)$ manifesting different power laws at different band fillings, as can be clearly seen in Fig. 1d  as well as Fig. S2c of Ref.~\onlinecite{jaoui2021quantum}.  This is in fact to be expected since, at low temperatures for $T<T^*$, the phonon scattering gets quenched and phonon-induced resistivity changes from the linear-in-$T$ behavior to a $T^4$ behavior, decreasing by 4 orders of magnitude and becoming essentially unobservable.  In this situation, many different scattering mechanisms come into play, including phonon drag,  impurity scattering, disorder scattering by twist angle fluctuations, electron-electron Baber and/or umklapp scattering, and possible scattering by the quantum fluctuations associated perhaps with any quantum critical point dominating the $T=0$ quantum phase diagram. We comment that scattering by spin or isospin (e.g. valley) fluctuations lead to a rather weak resistivity, which manifests a power law higher than linear, thus making these mechanisms unlikely to be the underlying resistive scattering mechanisms. A possible way to distinguish experimentally between quantum criticality and phonon scattering mechanisms for tBLG transport is to directly demonstrate the existence of critical fluctuations by measuring a diverging susceptibility, which to the best of our knowledge, has not yet been established in tBLG systems.  We do not anticipate any universal temperature dependence in this regime because of the presence of multiple different scattering processes competing with each other.  Ref.~\onlinecite{jaoui2021quantum} reports resistivity power laws on $T$ varying from 1 to 2, depending on the band filling, in the $T<1$ K regime.  It is unlikely that any linear-in-$T$ resistivity in the $T<1$ K regime could arise from the electron-phonon scattering mechanism  of interest in the current work, since according to Fig.~\ref{fig:1}, the crossover temperature scale for phonon scattering to manifest a linear-in-$T$ resistivity is $\sim 1$ K  even for a band filling as low as $\nu=0.1$.   We do not therefore have any explanation based on our minimal phonon scattering theory for any linear-in-$T$ behavior arising in the $T<1$ K regime in Ref.~\onlinecite{jaoui2021quantum}.  We do, however, mention that low temperature ($<10$ K) resistivity of even regular normal metals is not explicable based on any single scattering mechanism, and typically one must combine several different scattering sources in a rather arbitrary manner to make sense of the low-$T$ resistivity of metals in sharp contrast to the high-$T$ ($>40$ K for metals, $>10$ K for tBLG) linear-in-$T$ resistivity which is generic and universal, and is caused by phonon scattering.~\cite{macdonald1950resistivity}

There are known situations in semiconductor based 2D systems ~\cite{Lewalle2002PRB,Noh2003PRB,Lilly2003PRL,Manfra2007PRL,sarma2014PRB}, where an approximate low-temperature linear-$T$ resistivity may manifest in dilute carrier systems, arising from the non-phonon mechanism of the interplay between disorder and screening effects~\cite{sarma2015screening}, but this mechanism is very unlikely to play any role in graphene or tBLG.  Also, a Hubbard-type strongly correlated model may produce a linear-in-$T$ resistivity from umklapp electron-electron scattering at very `high' temperatures, but this is essentially a classical effect of energy equipartition with absolutely nothing `strange' in this linearity.  It is unlikely that the linear-in-$T$ resistivity behavior observed in tBLG \cite{jaoui2021quantum,Polshyn2019,Cao2020PRL} has anything to do with the high-$T$ Hubbard model properties.  In fact, we believe that any umklapp electron-electron interaction is likely to produce a $T^2$ low-temperature resistivity in tBLG  itself as has recently been argued in the literature.\cite{wallbank2019excess,ishizuka2021wide}  Thus, the appearance of a linear-in-$T$ resistivity in tBLG down to arbitrarily low temperatures for some specific band fillings, as reported in Ref.~\onlinecite{jaoui2021quantum} (and earlier in Ref.~\onlinecite{Cao2020PRL}), remains a mystery at this stage, but the observed higher-$T$ linear-in-$T$ resistivity most likely arises from the phonon scattering enhanced strongly by flat band moir\'e effects.

Before concluding we mention that a pronounced linear-in-$T$ resistivity has also recently been reported \cite{chu2021phonons,Tomic2022} in moir\'e twisted double bilayer graphene (tDBLG) as predicted earlier in \cite{Li2020PRB} based on phonon scattering considerations, with a strong flat band induced enhancement of the temperature coefficient of the resistivity with respect to the untwisted regular DBLG.\cite{Min2011PRB}  Thus, the phenomenon of phonon-induced pronounced strange metallic behavior may be a generic property of all moir\'e systems where the carrier Fermi velocity is strongly suppressed by the moir\'e band structure.  We should also comment on the so-called Planckian behavior of the phonon physics.~\cite{Cao2020PRL,Senthil2021,greene2020strange,Zaamen2019,Cha2021PRL,hartnoll2021planckian}  Converting the electron-phonon deformation potential $D$ into an effective dimensionless coupling constant $\lambda$, we get $\lambda \sim 0.5 -1$ for tBLG \cite{wu2019phonon}, which should be contrasted with a $\lambda \sim 0.0001$ for regular untwisted MLG.\cite{Hwang2008PRB}  Thus, the scattering rate, $1/\tau$, in tBLG for small twist angles, $\hbar/\tau = 2 \pi \lambda  k_B T$, is roughly 3-6 times the temperature (similar to the situation in strong electron-phonon  coupling metallic systems such as Pb), making tBLG a super-Planckian metal, strongly violating the Planckian bound of $\hbar/\tau < k_B T$, as has been discussed in the literature.\cite{jaoui2021quantum,Cao2020PRL,hartnoll2021planckian}  The super-Planckian behavior in our theory is of course neither strange nor mysterious (for $T>10$ K at least) since it arises from enhanced phonon scattering under the moir\'e flat band conditions.

To conclude, acoustic phonon scattering provides a reasonable generic explanation for the observed temperature dependence of the tBLG resistivity reported in Ref.~\onlinecite{jaoui2021quantum}, but cannot explain the transport properties for $T<1 \text{ K}$.  More work would be necessary to figure out the scattering mechanisms in tBLG at low temperatures below $1 \text{ K}$ where phonon scattering is likely strongly suppressed because of low thermal phonon occupancy in the Bloch--Gr\"uneisen regime.

\section{ ACKNOWLEDGMENTS }
We acknowledge helpful communications with Alexandre Jaoui and Dima Efetov. We are particularly grateful to Alexandre Jaoui for providing us the temperature coefficient of the experimental resistivity presented in Ref.~\onlinecite{jaoui2021quantum} for $T>10$ K as a direct data set facilitating an accurate comparison between theory and experiment. This work is supported by the Laboratory for Physical Sciences.  F. Wu is also supported by National Key R$\&$D Program of China 2021YFA1401300 and start-up funding of Wuhan University.

\bibliography{refs}
\end{document}